%% file: main.tex
\documentclass{article}

\usepackage{graphicx} 
\usepackage{amsmath,amsfonts,amssymb}
\usepackage{fullpage}
\usepackage{url}
\usepackage{todonotes}

\newtheorem{definition}{Definition}
\newtheorem{theorem}{Theorem}

\newtheorem{proposition}{Proposition}

\usepackage{xcolor}

\title{A Vector Representation for Phylogenetic Trees}
\author{Cedric Chauve$^{1}$,  Caroline Colijn$^{1}$,  Louxin Zhang$^{2}$\\
\\
\\
$^{1}$ Department of Mathematics\\ Simon Fraser University\\
Burnaby, BC V5A 1S6, Canada
\\
\texttt{[cedric.chauve,caroline\_colijn]@sfu.ca}\\
\\
$^2$ Department of Mathematics\\ National University of Singapore\\ Singapore 119076
\\
\texttt{mathzlx@nsu.edu.sg}
}
\date{}

\begin{document}

\maketitle

\begin{abstract}
\input{abstract.tex}
\end{abstract}

\vfill\pagebreak
\input{introduction.tex}

\vfill\pagebreak
\input{preliminaries.tex}

\vfill\pagebreak
\input{results.tex}

\vfill\pagebreak
\input{experiments.tex}

\vfill\pagebreak
\input{conclusion.tex}

\vfill\pagebreak
\bibliographystyle{RS}
\bibliography{reference}

\vfill\pagebreak

\input{appendix.tex}

\end{document}

%% file: abstract.tex
Good representations for phylogenetic trees and networks are important for optimizing storage efficiency and implementation of scalable methods for the inference and analysis of evolutionary trees for genes, genomes and species. 
We introduce a new representation for rooted phylogenetic trees that encodes a binary tree on $n$ taxa as a vector of length $2n$ in which each taxon appears exactly twice. 
Using this new tree representation, we introduce a novel tree rearrangement operator, called a \textit{HOP}, that results in a tree space of diameter $n$ and a quadratic neighbourhood size.
We also introduce a novel metric, the \textit{HOP distance}, which is the minimum number of HOPs to transform a tree into another tree.
The HOP distance can be computed in near-linear time, a rare instance of a tree rearrangement distance that is tractable.
Our experiments show that the HOP distance is better correlated to the Subtree-Prune-and-Regraft distance than the widely used Robinson-Foulds distance.
We also describe how the novel tree representation we introduce can be further generalized to tree-child networks.

%% file: introduction.tex
\section{Introduction}
\label{sec:introduction}

Phylogenetic trees and networks are essential mathematical frameworks for modeling the evolution of biological entities~\cite{felsenstein2004inferring}. 
These structures have many applications, facilitating the study of gene, genome, and species evolution~\cite{dayhoff1975evolution,murphy2005dynamics},
conservation and biodiversity research~\cite{redding2008evolutionarily}, mapping human migration patterns~\cite{cann1987mitochondrial,ingman2000mitochondrial},  and investigations into the transmission dynamics of infectious diseases such as HIV, COVID-19, and influenza~\cite{boni2020evolutionary,sharp2010evolution}. 

Despite their utility, inferring phylogenetic trees presents significant computational challenges, with NP-hard time complexity arising under both parsimony-based and stochastic models~\cite{foulds1982steiner,roch2006}.
Commonly employed software tools for tackling the tree inference problem, such as RAxML~\cite{stamatakis2005raxml}, IQ-TREE~\cite{minh2020} or BEAST~\cite{suchard2018bayesian}, typically navigate the vast tree space by evaluating millions of phylogenetic trees to maximize their optimization criteria. 
Consequently, there is a growing demand for concise and informative tree representations to expedite processing, optimize storage efficiency, enable uniform sampling of trees, and facilitate the application of deep learning methods for tree inference~\cite{voznica2022deep}. 

The Newick format~\cite{felsenstein2004inferring} is the most widely-used tree representation for computer input/output interface in bioinformatics. 
In Newick format, the topology of a phylogenetic tree is represented as an arrangement of parentheses, taxa labels and commas.  
This representation is intuitive and is based on the well-known bijection between balanced parenthesis words and ordered rooted trees. 
Generally, the Newick representation is not adapted to the manipulation of trees (e.g. performing tree rearrangements, deciding if two trees are identical) and merely serves as a way to save trees on disk. 

A few other representations have been investigated for phylogenetic trees~\cite{kim2020distance,penn2023phylo2vec,voznica2022deep}, labeled trees and tree shapes~\cite{liu2022analyzing}.
For instance, the Pr\"{u}fer code is a bijective encoding of a labeled tree with $n$ vertices using a sequence of $n-2$ integers between $0$ and $n-1$~\cite{prufer1918neuer}. 
A binary phylogenetic tree is uniquely represented as a $n-1$ dimensional vector of  integers $(a_1, a_2, \cdots, a_{n-1})$, where  $0\leq a_i\leq 2(j-1)$~\cite{rohlf1983numbering}.
This approach leverages the established fact that any phylogenetic tree on $n$ taxa can be derived from a single-leaf tree by sequentially attaching the remaining $n-1$ taxa~\cite{felsenstein2004inferring}. 
Termed Phylo2Vec, this vector presentation was recently investigated by Penn et al. for its utility in tree sampling for maximum likelihood tree inference~\cite{penn2023phylo2vec} among other applications.

Tree rearrangement operators are necessary for traversing the phylogenetic tree space (for example in Bayesian phylogenetics) and random tree sampling. 
Widely used tree rearrangement operators include  the Nearest-Neighbour-Interchange (NNI)~\cite{moore1973iterative} and the Subtree-Prune-Regraft (SPR). 
An SPR removes a subtree from  a tree and then re-grafts it on a branch of the resmaining tree. 
An NNI is a special case of an SPR in which  the branch onto which the pruned subtree is regrafted and the branch where the subtree was pruned in the original tree are next to each other. 
Any two phylogenetic trees on the same taxa can be converted into each other using a sequence of SPR (resp. NNI) rearrangements. 
The SPR distance between two trees is the minimum number of SPR rearrangements that are necessary to convert one into the other. 
The NNI distance between two trees is defined similarly.  
Both the NNI and SPR distance are tree metrics whose computation is intractable (i.e. NP-complete~\cite{dasgupta1997distance}), but in some special cases (such as the NNI distance fr ranked trees~\cite{colienne2021computing}).
Another widely-used tree metric is the Robinson-Foulds distance that is defined as the number of clades appearing in one tree but not in the other~\cite{robinson1981}.

Here, we present a vector representation for phylogenetic trees, derived from a decomposition technique for trees proposed in~\cite{zhang2023fast}.
In addition, the vector representation can be generalized to the family of tree-child phylogenetic networks \cite{cardona2008comparison}.
Using this representation, we introduce a new tree rearrangement operator and a related distance metric for tree comparison that, unlike most tree rearrangement based distances, is computable in polynomial time. 
Our experiments show that the vector representation we introduce allows to write trees using less disk space than the classical Newick format, and that the new distance we introduces approximates better the SPR difference between trees than the widely used Robinson-Foulds distance.

%% file: preliminaries.tex
\section{Preliminaries}
\label{sec:preliminarise}

\paragraph{Phylogenetic trees.}
Phylogenetic trees are graphic representations of the evolutionary relationships among biological entities (e.g. species, groups, or genes). 
In this paper, we define a \textit{phylogenetic tree} on a set $X$ of $n$ taxa as a rooted directed tree such that (i) the root is of outdegree $1$ and indegree $0$; (ii) all nodes other than the root are of indegree $1$; (iii) the nodes of outdegree $0$ are called the leaves that are in one-to-one correspondence with the $n$ taxa (Figure~\ref{fig2_encode}), (iv) all non-leaf and non-root nodes have outdegree $2$ (a phylogenetic tree is a binary tree). 
We denote by $V(T)$ the set of nodes of tree $T$.

Note that, for technical convenience, the definition above adds an extra node, of outdegree $1$, at the root of the tree. 
The nodes other than the root and the leaves are called \textit{internal nodes}. 
A phylogenetic tree is \textit{binary} if its internal nodes are of outdegree $2$.
Consequently, a binary phylogenetic tree on $n$ taxa has one root, $n-1$ internal nodes and $n$ leaves. 

Let $T$ be a phylogenetic tree on taxa $X$.  
We use $V(T)$ and $E(T)$ to denote the set of nodes and edges of $T$, respectively.   
For any $u, v\in V(T)$, $u$ is the \textit{parent} of $v$  (and $v$ is a \textit{child} of $u$) if $(u, v)\in E(T)$. 
The node $u$ is an \textit{ancestor} of $v$ if there is a directed path from $u$ to $v$. 
In this work, we simply say that $u$ is above $v$ and $v$ is below $u$ if $u$ is an ancestor of $v$.

We denote by $L(T)$ the set of leaves of $T$.
For $u\in V(T)$, we use $c(u)$ to denote the set of children of $u$ and $L(u)$ to denote the set of taxa (i.e leaves) below $u$ and call it the \textit{clade} at the node $u$. Clades are also called \textit{node clusters}.
For $u\in V(T)$, we denote by $\min(u)$ the smallest taxon of the clade $L(u)$.

In this work, a binary phylogenetic tree is simply called a tree. 
We consider leaf-labeled trees, i.e. trees with labels only on the leaves (internal nodes are not labelled) and where no two leaves have the same label.

\paragraph{Phylogenetic networks.}

A phylogenetic network on taxa $X$ is a binary rooted acyclic directed graph in which 
(i) the root is of outdegree $1$ and indegree $0$; (ii) the nodes of outdegree $0$ are called the leaves and are in one-to-one correspondence with the taxa; (iii) all the nodes other than the root  and leaves are of indegree $1$ and outdegree $2$, or indegree $2$ and outdegree $1$. 
The nodes with indegree $2$ and outdegree $1$ are called the \textit{reticulate nodes}. 
The nodes with indegree $1$ and outdegree $2$ are called the \textit{tree nodes}. 

For a phylogenetic network $N$ on $X$, we use $V(N)$ and $E(N)$ to represent the set of nodes and edges of $N$, respectively. 
The set of reticulate nodes are denoted with $R(N)$. 
Clearly, $V(N)\setminus R(N)$ consists of tree nodes and leaves of $N$.
A phylogenetic network is a tree-child network if every non-leaf node has at least one child that is not a reticulation.

\paragraph{Vectors and Longest Common Subsequence.}
Let $\Sigma$ be an alphabet. 
A \textit{vector} of length $n$ on $\Sigma$ is an ordered sequence $(v_1,v_2,\dots,v_n)$ of elements of $\Sigma$.  
The \textit{length} of  a vector $s$ is denoted by $\vert s\vert$.
It is a \textit{permutation vector} if each element of $\Sigma$ occurs exactly once in the vector. 
A \textit{partial permutation vector} is a vector satisfying the constraint that each element of $\Sigma$ occurs at most once in the vector.  
We use $\epsilon$ to denote the empty vector. 

A vector is a \textit{subsequence} of another if the former can be obtained from the latter by the deletion of zero or more entries. 
A vector is a \textit{common subsequence} of multiple vectors  if it is a subsequence of each vector. 

Let $S$ be a set of vectors. 
A  vectors is the \textit{Longest Common Subsequence} (LCS) for the vectors $S$ if it is (i) a subsequence of each vector in $S$ and (2) among all such common subsequences it has the largest length.
There may be multiple LCSs for the vectors $S$ and we denote by ${\rm LCS}(S)$  the set of all LCSs of $S$ and by $\vert {\rm LCS}(S)\vert$ their length.

%% file: results.tex
\section{Results}
\label{sec:results}

In this section we first introduce the novel representation of trees as vectors, together with linear-time algorithms to encode a tree into a vector and decode a vector into a tree. 
We then introduce the hop tree rearrangement operator and properties of the tree space it defines.

\subsection{Vector representation of rooted phylogenetic trees}
\label{ssec:representation}

Our representation for phylogenetic trees arises from  a path decomposition technique proposed in~\cite{zhang2023fast}.
We assume without loss of generality that the set of taxa of a given tree is $X=\{1,2,\dots,n\}$.

\begin{definition}
    \label{def:representation}
    A vector $v$ on the alphabet $X=\{1,2,\dots,n\}$ is a \textit{tree representation} if 
    \begin{enumerate}
        \item $\forall i \in X$, $i$ appears exactly twice in $v$;
        \item $v_1 = 1$;
        \item $\forall i \in X-\{1\}$, the first occurrence of $i$ in $v$ appears before the second occurrence of $i-1$;
        \item $\forall i \in X-\{1\}$, the second occurrence of $i$ in $v$ appears after the second occurrence of $i-1$.
    \end{enumerate}
\end{definition}

\noindent\textbf{Remark.}
For clarity, we will use $\color{red}\underline{i}$ to represent the second occurrence of $i$ in the a vector.

\begin{figure}[htbp]
    \centering
    \includegraphics[scale=0.5]{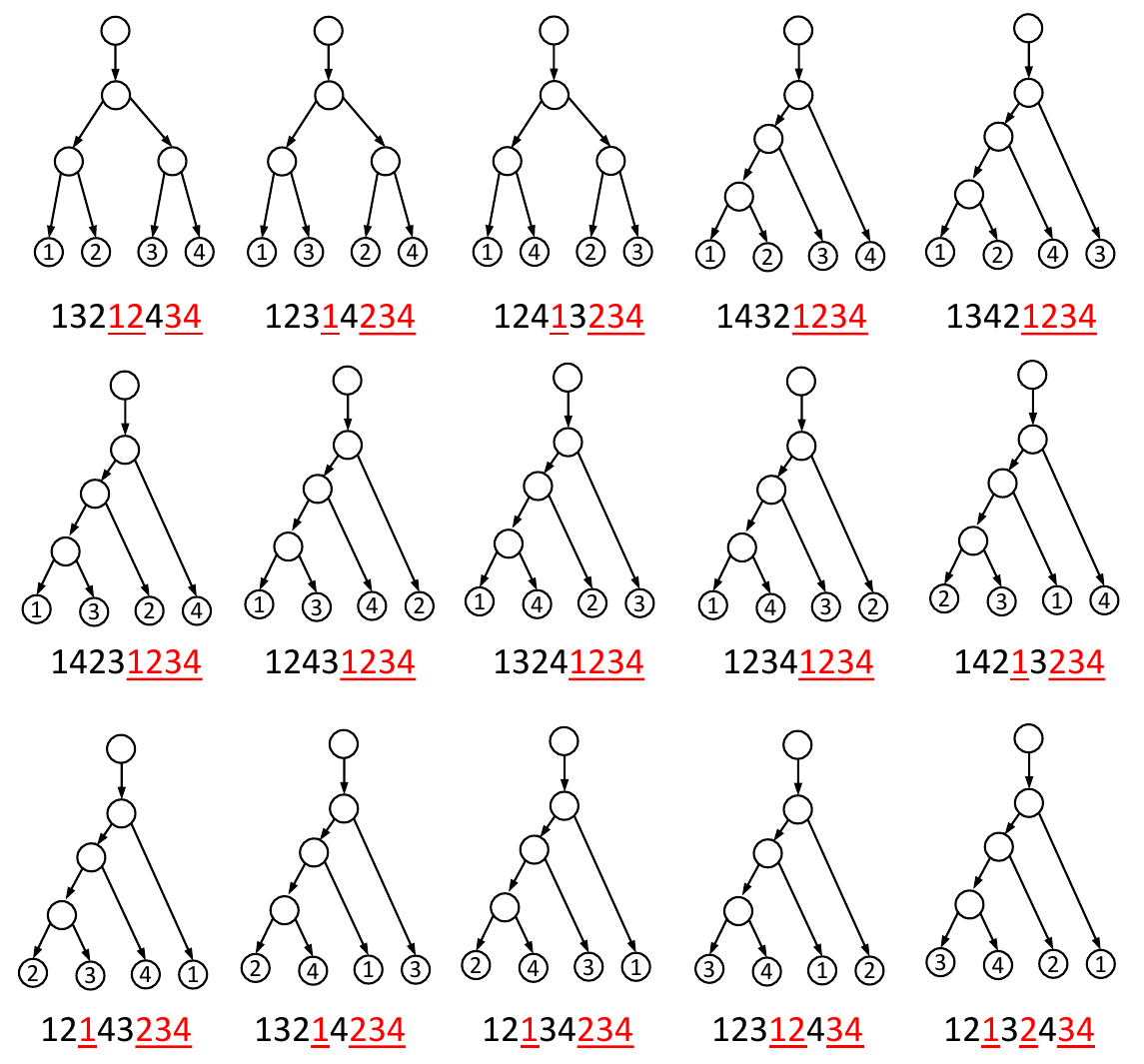}
    \caption{The fifteen trees on $X=\{1, 2, 3, 4\}$ and their representations. Here, commas are omitted in the tree representations, in which the second occurrences of the taxa are colored  and underlined.}
    \label{fig2_encode}
\end{figure}

\medskip
For example, $(1,3,2,{\color{red} \underline{1},\underline{2}},4,{\color{red}\underline{3},\underline{4}})$ is a tree representation for $n=4$, while $(1,3,{\color{red} \underline{1}},2,4,{\color{red} \underline{2},\underline{3},\underline{4}})$ is not, as the first occurrence of $2$ appears after the second occurrence of $1$.
There are 15 trees on four taxa, shown in Figure~\ref{fig2_encode} with their respective vector representations.

\begin{theorem}
    \label{thm:representation}
    There is a one-to-one correspondence between tree representations vectors on $X=\{1,2,\dots,n\}$ and rooted phylogenetic trees on $X$.
\end{theorem}

We describe algorithms to encode a tree into a tree representation and decoding a tree representation into a tree. 
It is straightforward to verify that the vector computed by the encoding algorithm satisfies Definition~\ref{def:representation} and that the decoding and encoding algorithm define a one-to-one correspondence between tree representations and trees.
Hence, Theorem~\ref{thm:representation} is proven. Recall that
$\min(u)$ denoted the smallest taxon of the clade $L(u)$ consisting of all descendant taxa of $u$.

\bigskip
\begin{tabular}{l}
    \hline 
    {\sc Tree Encoding Algorithm}\\
    \textbf{Input:}  A phylogenetic tree $T=(V, E)$\\
    \textbf{Output:} A tree representation $\mathbf{v}(T)$ \vspace{0.5em}\\
    1. For each $u \in V(T)$: compute $\min(u)$; \\
    2. For each $u \in V(T)-L(T)$: label $u$ by $\ell(u)=\max\limits_{v\in c(u)} \min(u)$;\\
    3. For $i\in X$: \\
    \hspace*{1.5em} 3.1. Let $P_i=(u_1,\dots,u_k)$ be the path from the non-leaf node labelled $i$ and the leaf $i$;\\
    \hspace*{1.5em} 3.2. Let $\mathbf{v}(P_i) = (\ell(u_2),\dots,\ell(u_{k-1}))$;\\
    4. Ouput 
    $\mathbf{v}(T)=(1,\mathbf{v}(P_1),1,\mathbf{v}(P_2),2,\cdots,\mathbf{v}(P_n),n)$;\\
    \hline
\end{tabular}

\bigskip
\begin{definition}
    \label{def:LTS}
    (\cite{zhang2023fast}) 
    The vector $\mathbf{v}(P_i)$ is called the labeling taxon sequence (LTS) of taxon $i$. 
\end{definition}

For the decoding algorithm, we compute from a tree representation on $X=\{1,2,\dots,n\}$ the set of edges $E(T)$ of a tree on $X$; to describe edges, we label non-leaf nodes by $\{n+1,\dots,2n\}$.
For a tree representation $\mathbf{v}=(v_1,\dots,v_{2n})$ and $v_i=x\in X$, we denote by $o(i)=1$ (resp. $o(i)=2$) if $v_i$ is the first (resp. second) occurrence of $x$ in $v$.

\bigskip
\begin{tabular}{l}
    \hline
    {\sc Decoding Algorithm}\\
    \textbf{Input:} A tree representation $(v_1,v_2,\cdots,v_{2n})$, $v_i\in X=\{1, 2, \cdots, n\}$ \\
    \textbf{Output:} The edge set $E(T)$ of a tree $T$\vspace{0.5em}\\
    1. Compute $o$ s.t. $o(i)=1$ (resp. $o(i)=2$) if $v_i(=x)$ is the first (resp. second) occurrence of $x$ in $v$;\\
    2. For $j$ from $1$ to $2n-1$:\\
    \hspace*{1.5em} 2.1. If $o({j})=1$ and $o({j+1})=1$: add $(n+v_j, n+v_{j+1})$ to $E(T)$;\\
    \hspace*{1.5em} 2.2. If $o({j})=1$ and $o({j+1})=2$: add $(n+v_j, v_{j+1})$ to $E(T)$;\\
    \hspace*{1.5em} 2.3. If $o(j)=2$ and $o({j+1})=1$:\\
    \hspace*{3.5em} 2.3.1. Let $j' = \min\limits_{k>j+1,o(k)=2} k$ (index of the first taxon whose second occurrence appears after $v_{j+1}$);\\
    \hspace*{3.5em} 2.3.2.  Add $(n+v_{j'}, n+v_{j+1})$ to $E(T)$;\\
    \hspace*{1.5em} 2.4. If $o(j)=2$ and $o({j+1})=2$: add $(n+v_{j+1}, v_{j+1})$ to $E(T)$;\\
    3. Output $E(T)$;\\
    \hline
\end{tabular}

\begin{figure}[htbp]
    \centering
\includegraphics[scale=0.6]{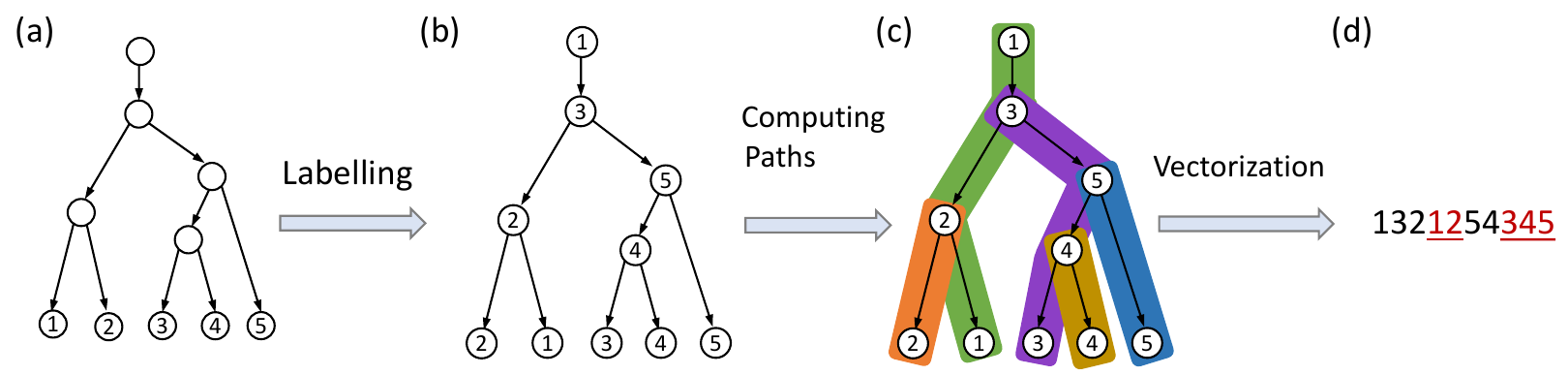}
    \caption{An illustration of the encoding of a phylogenetic tree into a tree representation. (a) A phylogenetic tree $T$ on $X=\{1, 2, 3, 4, 5\}$. (b) The labeling of internal nodes (step 2 of the encoding algorithm). (c) The  decomposition of the tree into the paths from the non-leaf node labelled $i$ to the leaf $i$. (d) The tree representation $\mathbf{v}(T)$, where the second copy of each taxon is underlined.}
    \label{fig1}
\end{figure} 

\bigskip
\noindent\textbf{Example.} Figure~\ref{fig1} provides an example of encoding of a tree into a tree representation.
Conversely, 
The table below shows the edges computed using the decoding algorithm applied to the tree representation  $(1,3,2,{\color{red} \underline{1},\underline{2}},5,4, {\color{red} \underline{3},\underline{4},\underline{5}})$ on $X=\{1,2,3,4,5\}$ ($n=5$).

\begin{center}
    \begin{tabular}{c|c|c|c}
        \hline
        $j$ & $v_j,v_{j+1}$ & $o(v_j),o(v_j+1)$ & edge \\
        \hline
        $1$ & $1,3$ & $1,1$ & $(6,8)$\\
        $2$ & $3,2$ & $1,1$ & $(8,7)$\\
        $3$ & $2,\underline{1}$ & $1,2$ & $(7,1)$\\
        $4$ & $\underline{1},\underline{2}$ & $2,2$ & $(7,2)$\\
        $5$ & $\underline{2},5$ & $2,1$ & $(8,10)$ ($s_{j'}=3$) \\
        $6$ & $5,4$ & $1,1$ & $(10,9)$\\
        $7$ & $4,\underline{3}$ & $1,2$ & $(9,3)$\\
        $8$ & $\underline{3},\underline{4}$ & $2,2$ & $(9,4)$\\
        $9$ & $\underline{4},\underline{5}$ & $2,2$ & $(10,5)$\\
        \hline
    \end{tabular}
\end{center}

\noindent One can verify that, up to the labels of the internal nodes, the tree defined by this set of edges is identical to the tree shown in Fig.~\ref{fig1}a.

\bigskip
\noindent \textbf{Remarks.} 
\begin{enumerate}
    \item The tree representation can also be used for unrooted phylogenetic trees by encoding taxa as $0,1,\cdots,n$ and rooting the tree at the leaf $0$. 
    \item The encoding and decoding algorithms can be implemented with a linear $(\Theta(n))$ time and space complexity.
    \item If $T$ is a tree with taxa that are not labeled by $X = \{1,\dots,n\}$, the results described above apply if an indexing $\sigma$ of the leaves of $T$ is provided that allows to map the taxa to $X=\{1,\dots,n\}$.
    As a consequence, different taxa indexings can lead to different vectors for the same tree.
    However, for a fixed indexing of the taxa, for any two trees $T_1 \neq T_2$, $\mathbf{v}(T_1)\neq\mathbf{v}(T_2)$ and Theorem~\ref{thm:representation} applies.
    \item In the representation $\mathbf{v}(T)$, the first occurrence of each taxon represents an internal node. 
    The second occurrence of  each $i$ is used to separate the LTSs $\mathbf{v}(P_i)$, $1\leq i\leq n$ and represents he leaf labelled $i$.
    \item Given a tree whose nodes (both internal nodes and leaves) are named and with branch lengths associated to edges, the vector representation of $T$ can be augmented to encode the full information of the tree as follows: every entry $v_i=x$ is replaced by a triplet $v_i=(x,y,\ell)$ where $y$ is the name of the corresponding node and $\ell$ the length of the branch from $y$ to its parent.
    The taxa order used for the encoding can then be deduced from the order of the entries of $\mathbf{v}$ corresponding to leaves.
    \item From a practical point of view, the vector encoding described above can be used to save trees on disk in a slightly more efficient way than the widely use Newick format.  
    We refer Section~\ref{sec:results} for an illustration of this claim. 
    To provide an example, the two trees, described in Newick format with internal node names and branch lengths
\begin{verbatim}
((b:1.0,d:2.0)e:2.0,(a:1.0,c:1.0)g:1.0)f;
(((b:1.0,d:2.0)x:2.0,a:1.0)y:1.0,c:1.0)z;
\end{verbatim}
    can be written, assuming their encoding uses the taxa order \texttt{a,b,c,d}, as
\begin{verbatim}
#a,b,c,d
1::0.0,2:f:0.0,3:g:1.0,1.0,4:e:2.0,1.0,1.0,2.0;
1::0.0,3:z:0.0,2:y:1.0,1.0,4:x:2.0,1.0,1.0,2.0;
\end{verbatim}
    
\end{enumerate}

\subsection{The HOP tree rearrangement operator}
\label{ssec:HOP}

The vector encoding we introduced in the previous section allows to define a novel tree rearrangement operator, together with a polynomial-time computable novel metric on trees.

\begin{figure}[htbp]
    \centering
    \includegraphics[scale=0.5]{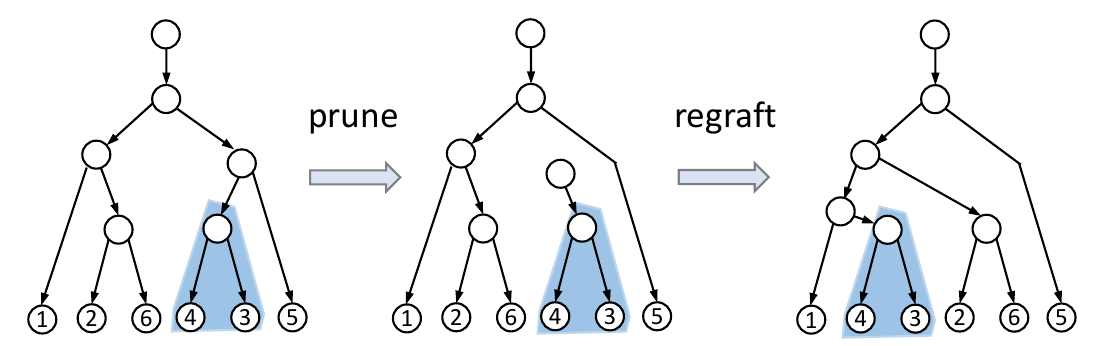}
    \caption{An illustration of the SPR operator.}
    \label{fig:fig_spr}
\end{figure}

Widely tree rearrangement operators used for traversing the space of phylogenetic trees include the Nearest-Neighbour Interchange (NNI) and the Subtree-Prune-and-Regraft (SPR), illustrated in  Fig.~\ref{fig:fig_spr}.
Recall that the NNI  rearrangement is a special type of SPR, where the branch of the pruned subtree and the branch where it is regrafted are co-localized in the tree.

The space of phylogenetic trees can be explored by moving from a tree $T$ to a tree that differs from $T$ by a single rearrangement; the set of such trees is the \textit{neighbourhood} of $T$. 
The size of the NNI-neighbourhood of a tree $T$ contains $\Theta(n)$ trees, while the space of the SPR-neighbourhood contains $\Theta(n^2)$ trees.
The diameter of the tree space defined by the NNI operator is bounded abov by $O(n\log_2(n))$ while the diameter of the space defined by the SPR operator is $n-\theta(\sqrt{n}))$~\cite{stjohn2016review}.
Any tree can be converted into another using a maximum of $(n-3)$ SPR rearrangements~\cite{allen2001subtree}.
However, there are two trees such that are at least $\frac{1}{4}n\log n-o(n)$ NNI away from each other~\cite{li1996nearest}.

Here,  we introduce a new tree rearrangement operator, that is a special type of SPR, that we call the \textit{HOP}.
As we will see, 
the HOP operator defines a tree space of diameter $n$, with every tree having a neighbourhood of size $O(n^2)$. 

\begin{figure}[htbp]
    \centering
    \includegraphics[scale=0.5]{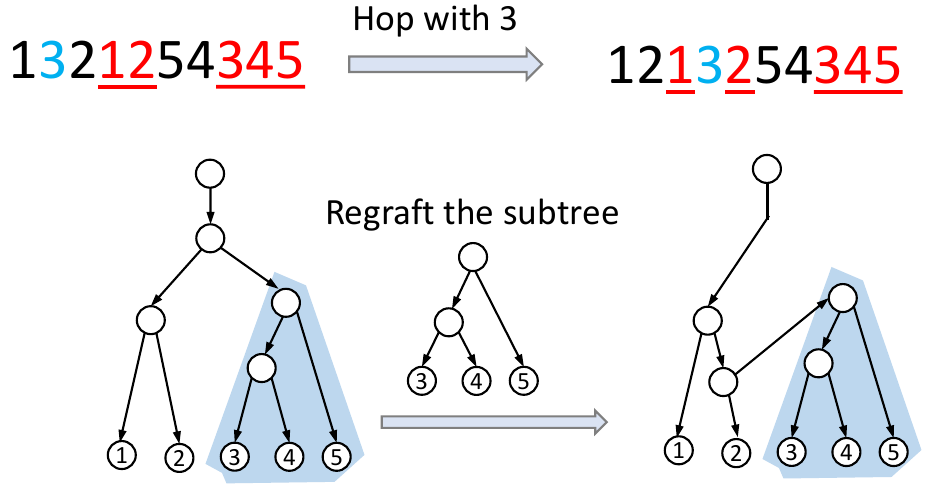}
    \caption{An illustration of the HOP operator that moves $3$ before $\underline{2}$, and its effect as an SPR rearrangement. The only other HOP rearrangement moving $3$ would move it before $\underline{1}$.}
    \label{fig:HOP}   
\end{figure}

\begin{definition}
  Let $\mathbf{v}$ be a vector representation of a phylogenetic tree on $X=\{1, 2, \cdots, n\}$. 
  We use $i$ and $\underline{i}$ to denote respectively the first and the second occurrence of $i$ in $\mathbf{v}$, respectively,  for each $1\in X$. 
  A \textit{HOP rearrangement} on $i\in X, i>1$ in $\mathbf{v}$ consists in moving $i$ to a new position located between $1$ and $\underline{i-1}$  in $\mathbf{v}$ (see Fig.~\ref{fig:HOP}). We denote by $(j,k)$ the HOP moving the element $v_j$ before the element in position $v_k$.
\end{definition}

\noindent
\bigskip\textbf{Remark.} The HOPs defined by $(j,j)$ and $(j,j+1)$ have no effect on $\mathbf{v}(T)$.

\begin{proposition}
    \label{prop:HOP_properties}
    \begin{enumerate}
        \item Any HOP rearrangement is an SPR rearrangement.
        \item For a given taxa ordering, the graph whose vertices are all tree representations for trees over the same set of taxa and edges connect any pair of trees that differ by a single HOP is connected.
        \item The size of the HOP neighbourhood of any tree $T$ on $n$ taxa is quadratic in $n$.
    \end{enumerate}
\end{proposition}
\textbf{Proof} We consider a tree $T$  on taxa $X=\{1, 2, \cdots, n\}$, with vector representation $\mathbf{v}(T)$. 

(1).  Recall that $\mathbf{v}(T)$ is derived from a decomposition of $T$ into $n$ paths $P_1, P_2, \cdots, P_n$, where $P_i$ starts from the internal node represented by the first occurrence of $i$ (denoted by $u$) in $\mathbf{v}(T)$ to the leaf $i$ (Fig.~\ref{fig1}). 
Let $u'$ be the child of $u$ on the path to the leaf $i$.
Removing $i$ from $\mathbf{v}(T)$ consists in pruning the subtree rooted at $u'$.
Inserting $i$ before $v_j$ consists in regrafting this subtree on the edge from the node corresponding to $v_j$ to its parent.
The constraint that $v_j$ must be not be after $\underline{i-1}$ in $\mathbf{v}(T)$ imposes that this regrafting occurs on a path $P_k$ for which $k\leq i-1$; by construction, none of these paths is in the subtree rooted at $u'$, so this regrafting is a valid SPR.

(2). It is straightforward to see that any tree can be transformed by a sequence of HOPs into the caterpillar tree whose vector representation is $(1,2,{\color{red}\underline{1}},3,{\color{red}\underline{2}},\dots,{\color{red}\underline{n-1},\underline{n}})$.
This proves the graph defined by the HOP operator over the space of tree representations for a given taxa ordering is connected.

(3). Let $p_k$ denote the position in $\mathbf{v}(T)$ of $\underline{k}$.
By definition, $p_k \geq 2k+1$ as all elements $j$ and $\underline{j}$, for $j=1,\dots,k-1$, must appear before $\underline{k}$, as must $k$ and $k+1$.
Moreover, for a given $i$, it can be moved by a HOP to any position before $p_{i-1}$, but before $1$ or at its current position. 
So the number of positions where $i \in\{2,\dots,n\}$ can be moved by a HOP is $p_{i-1}-2$.
This implies that the number of possible HOP rearrangements is at least
$\displaystyle \sum_{2\leq i\leq n} 2(i-1)-1$, which is quadratic in $n$.
Last, we need to account for overcounting, due to the fact hat two different HOP could result in the same tree. 
This can occur only for the following HOPs: $(j,j+2)$ and $(j+1,j)$ hat both swap the elements in positions $j$ and $j+1$.
So there is a only linear number of such pairs of equivalent HOPs.
This proves that the size of the HOP neighbourhood of any tree $T$ on $n$ taxa is quadratic in $n$.
If we denote by $\ell_i$ the boolean that is true if the element in position $i$ in $\mathbf{v}$ is the first occurrence of its label $v_i$ and false otherwise, the set of HOPs defining the neighbourhod of $\mathbf{v}$ can be described as
\begin{equation}
    \label{eq:HOP_ngb}
    \{ (j,k) \in X^2,
    \ j>1,
    \ \ell_{j} = \mathrm{true},
    \ 1 < k \leq p_{v_j-1},
    \ k \notin \{j,j+1,j+2\}
    \}.
\end{equation}

\bigskip
\noindent \textbf{Remarks.}
\begin{enumerate}
    \item The proof of Proposition~\ref{prop:HOP_properties}.(3) shows that it is easier with the HOP operator to define the neighbourhood than with the SPR operator where one has to consider if the tentative regrafting edge is not in the subtree rooted by the pruned edge.
    \item Proposition~\ref{prop:HOP_properties}.(2) implies that the HOP operator can be used to explore the space of all tree representations over a given taxa set.
    Equation~\ref{eq:HOP_ngb} shows that it is easy to uniformly chose a random HOP for a given tree representation $\mathbf{v}$.
    However, exploring the space of tree representations is different than exploring the space of trees, as the chosen taxa ordering implies a bias in the probability of subtrees pruning (the element $1$ in a tree representation is much less likely to be moved by a random HOP than the element $n$).
    This can be addressed by adding in the exploration process the possibility to switch from the current taxa ordering to a different (random) one.
\end{enumerate}

\subsection{The HOP distance for phylogenetic trees}
\label{ssec:HOP_dist}

Both NNI and SPR define a notion of distance between two trees, the minimum number of rearrangements to transform one tree into the other one, or equivalently the length of the shortest path between the two trees in the tree space defined by the operator. 
Both these distances are tree metrics and are NP-hard to compute exactly~\cite{stjohn2016review}.

Similarly to other tree rearrangements, we can define a distance based on the HOP rearrangement.

\begin{definition}
    \label{def:HOP_dist}
    Let $T_1$ and $T_2$ be two trees on taxa $X=\{1,\dots,n\}$, and $\mathbf{u}$ and $\mathbf{v}$ their respective vector representations. 
    The HOP distance $d_{\rm HOP}(\mathbf{u},\mathbf{v})$ is the minimum number of HOP necessary to transform $\mathbf{u}$ into $\mathbf{v}$.
\end{definition}

In this section we show that the HOP distance is a tree metric and can be computed in $O(n\log_2(n))$, through a variation of the LCS. 

\begin{definition}
    \label{def:HOP_sim}
    Let $T_1$ and $T_2$ be two trees on taxa $X=\{1,\dots,n\}$, and $\mathbf{u}=(\mathbf{u}_1,\underline{1},\mathbf{u}_2,\underline{2},\cdots,\mathbf{u}_{n-1},\underline{n-1},\underline{n})$ and \\
    $\mathbf{v}=(\mathbf{v}_1,\underline{1},\mathbf{v}_2,\underline{2},\cdots,\mathbf{v}_{n-1},\underline{n-1},\underline{n})$ their respective vector representations. 
    The HOP similarity is defined as
    $${\rm Sim}_{\rm HOP}(\mathbf{u},\mathbf{v})=\sum_{1\leq i\leq n-1} \vert {\rm LCS}(\mathbf{u}_i, \mathbf{v}_i)\vert.$$
\end{definition}

\begin{theorem}
    \label{thm:HOP_dist}
    Let $T_1$ and $T_2$ be two trees on taxa $X=\{1,\dots,n\}$, and $\mathbf{u}$ and $\mathbf{v}$ their respective vector representations. 
    $$d_{\rm HOP}(\mathbf{u},\mathbf{v}) = n-{\rm Sim}_{\rm HOP}(\mathbf{u},\mathbf{v}).$$
\end{theorem}

\noindent\textbf{Proof}
First $\mathbf{u}=\mathbf{v}$ if and only if ${\rm Sim}_{\rm HOP}(\mathbf{u},\mathbf{v})=n$.
Let $k = {\rm Sim}_{\rm HOP}(\mathbf{u},\mathbf{v})$.
A HOP operator moves a single element in a vector representation, which implies that it can increase the HOP similarity by at most $1$.
This implies that $d_{\rm HOP}(\mathbf{u},\mathbf{v}) \geq n-k$.
Next, it is easy to find a sequence of $n-k$ HOP that transform $\mathbf{u}$ into $\mathbf{v}$.
Let $v_i=x$ be the first element in $\mathbf{v}$ that is not in the LCS of the segment $\mathbf{v}_j$ it belongs to. 
The HOP rearrangement on $\mathbf{u}$ that moves $x$ immediately after $v_j$ does increase the HOP similarity by $1$.
This can be iterated until both vector representations are equal, in exactly $k$ HOP rearrangements.
This implies that $d_{\rm HOP}(\mathbf{u},\mathbf{v}) \leq n-k$.

\medskip
\begin{proposition}
    \label{prop:HOP_forest}
    The vector
    ${\rm LCS}(\mathbf{u},\mathbf{v})=({\rm LCS}(\mathbf{u}_1, \mathbf{v}_1),\underline{1},{\rm LCS}(\mathbf{u}_2, \mathbf{v}_2),\underline{2},\cdots, {\rm LCS}(\mathbf{u}_{n-1}, \mathbf{v}_{n-1}),\underline{n-1},\underline{n})$ defines a common forest shared by the trees  $T_1$ and $T_2$.
\end{proposition}
\noindent \textbf{Proof} 
Using the decoding algorithm, we can recover all the edges in the forest. 
In short, ${\rm LCS}(\mathbf{u}_i, \mathbf{v}_i)\underline{i}$ defines a path $P_i$ for each $i$. 
We obtain the forest by connecting the first occurrence of $i$ to the starting node of $P_i$ if the first occurrence of $i$ appears in the vector ${\rm LCS}(\mathbf{u}, \mathbf{v})$. 
Note that the start node of $P_i$ will be the root of a subtree if the first occurrence of $i$ does not appear in  ${\rm LCS}(\mathbf{u}, \mathbf{v})$.

\medskip
For example, consider the representations of the first two trees in Figure~\ref{fig2_encode}: 
$\mathbf{u}=(1,3,2,{\color{red} \underline{1},\underline{2}},4,{\color{red}\underline{3},\underline{4}})$ and $\mathbf{v}=(1,2,3,{\color{red}\underline{1}},4,{\color{red}\underline{2},\underline{3},\underline{4}})$. 
For these two representations, we have ${\rm LCS}(\mathbf{u},\mathbf{v})=(1,3,{\color{red}\underline{1},\underline{2},\underline{3},\underline{4}})$, which encodes the forest consisting of three subtrees $(1, 3), (2), (4)$ in Newick format.

\begin{proposition}
    \label{prop:HOP_properties2}
    \begin{enumerate}
    \item The HOP distance between two tree representations on $n$ taxa can be computed in near-linear time $O(n\log n)$.
  
    \item For a given taxa order, the diameter of the HOP tree space for trees on $n$ taxa is $n$.
    \item The HOP distance is a metric in the space of (rooted) phylogenetic trees under a fixed taxa order.
    \end{enumerate}
\end{proposition}

\noindent \textbf{Proof} (1) The result is derived from the fact that for each $i\leq n$, , if $n_i=\max(|\mathbf{u}_i|, |\mathbf{v}_i|)$, ${\rm LCS}(\mathbf{u}_i, \mathbf{v}_i)$ can be computed in  $O\left(n_i\log_2(n_i)\right)$ as  $\mathbf{u}_i$ and $\mathbf{v}_i$ are partial permutations of $X$, and the LCS between permutations can be reduced to the problem of computing a Longest Increasing Subsequence of a permutation of size $n$, which can be done in $O(n\log_2(n))$ (see \cite{hunt1977fast} for example).
\\

(2) By definition, the HOP similarity is non-negative. This implies that the HOP distance is at most $n$. 
\\

(3) For any two tree representations $\mathbf{u}$ and $\mathbf{v}$, by definition of the LCS, ${\rm SIM}_{\rm HOP}(\mathbf{u},\mathbf{v})={\rm SIM}_{\rm HOP}(\mathbf{v},\mathbf{u})\geq 0$ and thus $d_{\rm HOP}(\mathbf{u},\mathbf{v})=d_{\rm HOP}(\mathbf{v},\mathbf{u})$ and the distance is 0 if and only if $P=Q$.

 For any three tree vectors $\mathbf{u},\mathbf{v},\mathbf{w}$,  there is a sequence of $d_{\rm HOP}(\mathbf{u},\mathbf{v})$ HOP rearrangements that converts $\mathbf{u}$ into $\mathbf{v}$, and there is a sequence of $d_{\rm HOP}(\mathbf{v},\mathbf{w})$ HOP rearrangements that converts $\mathbf{v}$ to $\mathbf{w}$. 
 Combining these two sequences of HOP rearrangements, we obtain a sequence of $d_{\rm HOP}(\mathbf{u},\mathbf{v})+d_{\rm HOP}\mathbf{v},\mathbf{w})$ rearrangements that converts $\mathbf{u}$ to $\mathbf{w}$. 
 This implies the triangle inequality  
 $d_{\rm HOP}(\mathbf{u},\mathbf{w})\leq  d_{\rm HOP}(\mathbf{u},\mathbf{v})+d_{\rm HOP}(\mathbf{v},\mathbf{w})$.

\medskip
\noindent\textbf{Remarks.}
\begin{enumerate}
    \item A useful consequence of the ability to compute the HOP distance using the vector
    ${\rm LCS}(\mathbf{u},\mathbf{v})$ is that it allows to compute easily a sequence of vector representations representing a shortest sequence of HOPs transforming $\mathbf{u}$ into $\mathbf{v}$, as described in the proof of Theorem~\ref{thm:HOP_dist}. 
    \item An undesired property of the HOP distance is that the representation of a tree is depends a specific indexing of the taxa.  
    Consider two trees $T_1,T_2$ on the same set $X$ of taxa, and two different indexings $\sigma,\sigma'$ of $X$.
    Let $\mathbf{u},\mathbf{v}$ (resp. $\mathbf{u}',\mathbf{v}'$) be the vector representations of $T_1$ and $T_2$ respectively, under $\sigma$ (resp. $\sigma'$).
    Then, it can occur that ${\rm Sim}_{\rm HOP}(\mathbf{u},\mathbf{v}) \neq  {\rm Sim}_{\rm HOP}(\mathbf{u}',\mathbf{v}')$, and so $d_{\rm HOP}(\mathbf{u},\mathbf{v}) \neq d_{\rm HOP}(\mathbf{u}',\mathbf{v}')$.
    This motivates us to introduce the \textit{mHOP} distance, that we conjecture is hard to compute.
    \begin{definition}
        \label{def:HOP_generalized}
        The {mHOP} distance $d_{\rm mHOP} (T_1, T_2)$ between two trees $T_1$ and $T_2$ is equal to the minimum HOP distance between  the representations of $T_1$ and $T_2$ over all possible taxa orders.
    \end{definition}
\end{enumerate}

\subsection{Generalizations}

In this subsection, we outline how to obtain a vector representation for trees with polytomies and some phylogenetic networks.

\subsubsection{Trees with polytomies}
\label{sssec:polytomies}
Recall that to obtain the  representation of a binary tree, a node $u$ with children $u'$ and $u''$ is labeled with the larger of $\min(u')$ and $\min(u'')$.

In a phylogenetic tree, a node of outdegree more than two  represents a polytomy where the evolutionary relationships can not be fully resolved .  
To obtain a representation of a non-binary tree on taxa $X=\{1, 2, \cdots, n\}$, we label each internal node with outdegree two in the same way than for binary trees; we label a node $u$ with  children  $u_1, u_2, \cdots, u_k$ ($k>2$) with the subset $\{\min (u_1), \min (u_2), \cdots, \min (u_k)\} \setminus \{\min_{1\leq i\leq k}\min (u_i)\}$, which is the subset consisting of all $\min{u_i}$ except the smallest one. 
In this way, the representation of a non-binary tree can be encoded as a sequence of taxa, commas and braces, in which a pair of braces corresponds to a polytomy in the tree. 
For instance, $(1,3,\{24\},{\color{red}\underline{1},\underline{2},\underline{3},\underline{4})}$ is a representation of a tree with one polytomy. 

\subsubsection{Tree-child networks}
\label{sssec:netwoks}
The vector representation can be generalized to the family of \textit{tree-child networks}. 
A phylogenetic network is a tree-child network if every non-leaf node has at least one child that is not a reticulation, where a reticulation is a node with indegree two and outdegree one. 

\begin{figure}[t!]
    \centering
    \includegraphics[scale=0.4]{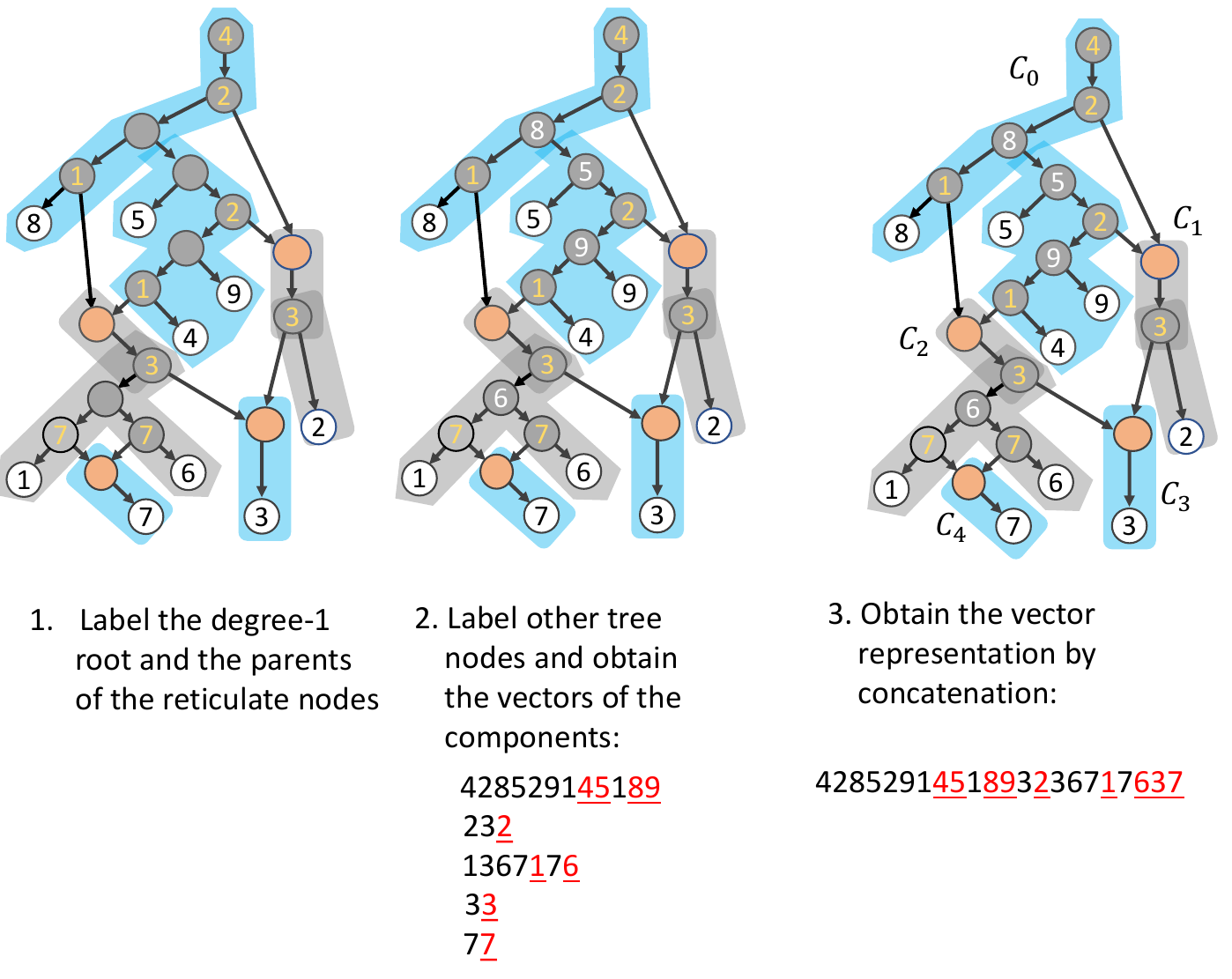}
    \caption{Illustration of the procedure of encoding tree-child networks.
    The five tree components of the network are highlighted in blue and grey. }
    \label{fig4}
\end{figure}

A connected component of the forest resulting from removing all the edges entering reticulated nodes from a tree-child network is called a tree component of the network (Fig.~\ref{fig4}).  
The root of a tree component is either the network root or a reticulation node. 
Consequently, a tree-child network with $k$ reticulation nodes has $k+1$ tree components. 
The tree component rooted at the network root is called the top component. 

To obtain the representation of a tree-child network $N$, we need to order the tree components (and thus the reticulation nodes) of $N$ such that (1) the top component is the first component in the order, and (2) a tree component $A$ appears before another tree component $B$ implies there is no directed path from any node in $B$ to any node in $A$.  
For example, we can order the tree components as follows.
\begin{itemize}
    \item Label the degree-1 root by the the smallest taxa in the top component. 
    Similarly, label the two parents of each reticulation node $y$  with the smallest taxa in the tree component rooted at $y$ (Figure~\ref{fig4}a).

    \item If a tree component of the network contains $t$ taxa, it contains $t-1$ degree-2 tree nodes. 
    We label these tree nodes using one of the remaining $t-1$ taxa  and obtain the vector representation of the component by using the same rule as in the binary tree case. 
    See Fig.~\ref{fig4}b where we use the increasing order to order the taxa).
    
    \item  We order the components inductively. 
    First, the top component receives index $0$. 
    Assume the $k$ components have been indexed as $C_0, C_1, \cdots, C_{k-1}$.  
    Now, consider all non-indexed components whose root (which is a reticulation node) has its two parents already in the indexed tree components. 
    If both parents are in the same component, say $K$, then one appears after the other one in the representation of $K$; the latter  is called the second parent. 
    If both parents are in different components, then the parent in the component with the higher index is called  the second parent. 
    Now order all the above chosen non-indexed components according to where  the second parents are located.
    If the second parent of the tree component A appears in $C_i$ and the second parent of the tree component appears in $C_j$ and $i<j$, $A$ has an index smaller than $B$. 
    If the second parent of $A$ and $B$ appears in the same tree component $C_i$, $A$ has a smaller index than $B$ if the second parent of $A$ appears earlier than  that of $B$ in the vector representation of $C_i$.
    Iteat this process until all path-components are indexed which will eventually happen because our networks are connected and acyclic. 
    This step is illustrated in Fig.~\ref{fig4}c.
\end{itemize}

Assume the $k+1$ components of the network are indexed as $C_0, C_1, \cdots, C_k$ and 
the vector representation of $C_i$ be $r_i$. 
The vector representation of the network is the concatenation of $r_0, r'_1, \cdots, r'_k$, where $r'_i$ is obtained by removing the first element of $r_i$ (Fig.~\ref{fig4}c).

%% file: experiments.tex
\section{Experiments}
\label{sec:experiments}

The encoding and decoding algorithm, as well as the HOP similarity algorithm were implemented in python and are freely available at \url{https://github.com/cchauve/CEDAR}.

\paragraph{Encoding phylogenetic trees.}
We generated $1,000$ random rooted phylogenetic trees, on $100$ taxa labelled \texttt{t1,\dots,t100}, with random branch lengths.
The encoding of the trees using the vector representation as described in Section~\ref{ssec:representation} requires slightly less disk space ($92\%$) than the Newick encoding.

\paragraph{HOP neighbourhood size.}
For each of the $1,000$ random trees, we computed the size of the HOP neighbourhood (the trees differing from the starting tree by exactly one HOP), based on a vector encoding obtained with taxa ordered according to a postorder traversal of the first tree.
The mean HOP neighbourhood size is $12,424$, with a $183$ standard deviation, the smallest (resp. largest) HOP neighbourhood containing $9,958$ (resp. $13,000$) trees.
This illustrates that despite the HOP rearrangement operator being a special type of SPR, it defines a neighbourhood of quadratic size.

\paragraph{HOP distance, Robinson-Foulds distance and SPR.}

In the last experiment, we selected the first $50$ random trees from the previous experiments, and  generated from each  $20$ trees obtained by applying $5,10,15,\dots,100$ random SPR rearrangements respectively.
We call the initial trees the \textit{focal trees} and the trees obtained by SPR rearrangements the \textit{SPR trees}.

For every pair $(T,T')$ where $T$ is a focal tree and $T'$ an SPR tree obtained from $T$, we computed:
\begin{itemize}
    \item the widely used Robinson-Foulds (RF) distance between $T$ and $T'$, normalized by dividing it by $200$;
    \item the HOP distances, normalized by dividing it by $100$.
\end{itemize}
To assess the impact of the taxon indexing required to generate and encoding and compute the HOP distance on the vector encoding, we generated $10$ random taxa orders and repeated the HOP distance computation for each order.

\begin{figure}[htbp]
    \centering
    \includegraphics[scale=0.5]{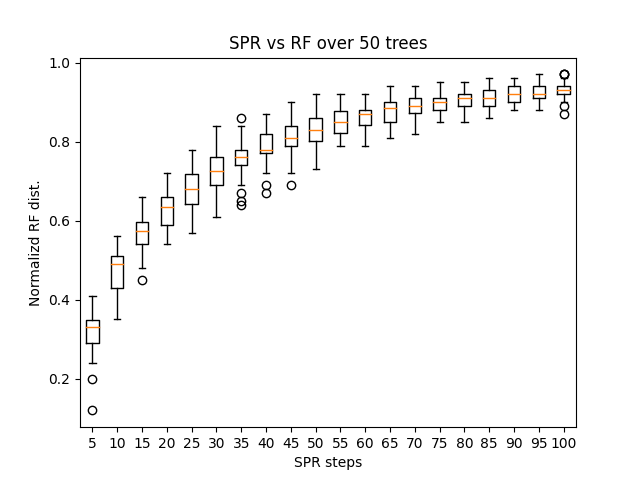}
    \includegraphics[scale=0.5]{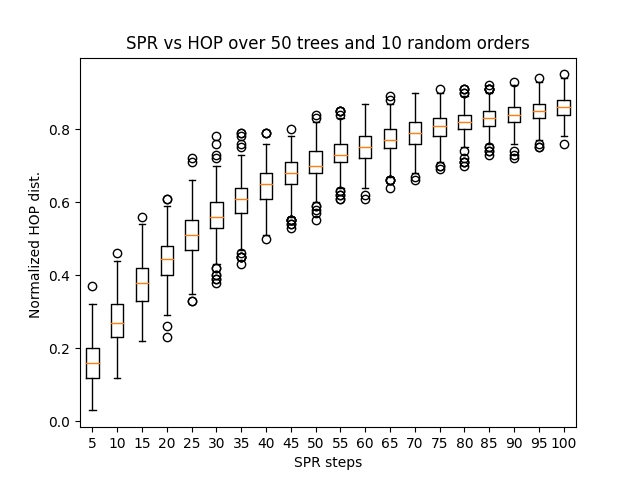}
    \caption{RF and HOP distances vs. the number of SPRs used to generate SPR trees from focal trees. The HOP distances are obtained from $50$ starting trees and $10$ random taxa order, so each box plot corresponds to $500$ distance values. Each box plot for the RF distance corresponds to 50 distance values.
    \label{fig:RF_vs_HOP}}
\end{figure}

Figure~\ref{fig:RF_vs_HOP} and Table~\ref{tab:correlation} shows that overall, the HOP distance correlates better with the actual number of SPRs used to generate SPR trees than the RF distance, especially for SPR trees obtained through a moderate number of SPR rearrangements.
However we observe a larger variation in the HOP distance value when considering 10 random taxa orders, compared to the RF distance, illustrating the effect of the chosen taxa order.

\begin{figure}[htbp]
    \centering
    \includegraphics[scale=0.5]{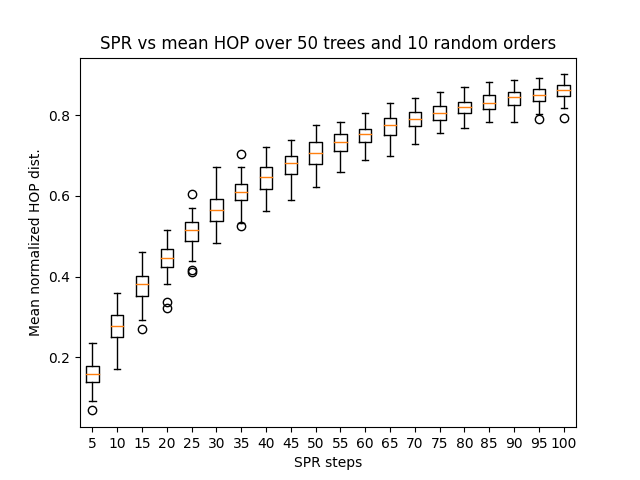}
    \includegraphics[scale=0.5]{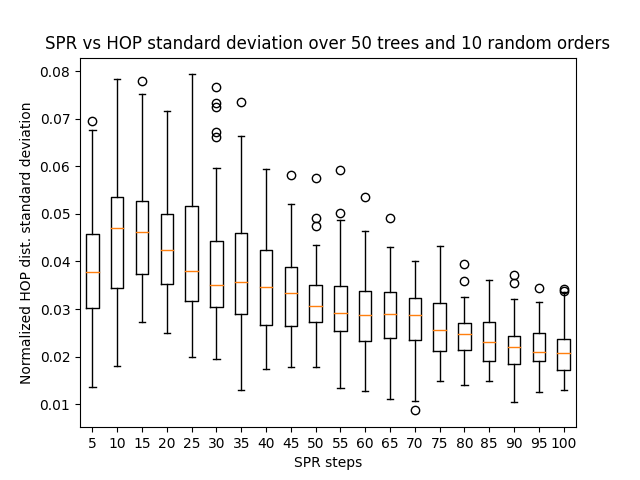}
    \caption{Mean and standard deviation of the HOP distance for $50$ starting trees, with mean distance and standard deviation over $10$ random taxa orders. Each boxplot shows the distibution of $50$ values. } 
    \label{fig:HOP_mean_std}
\end{figure}

Since the HOP distance depends implicitly on the taxa order used to generate vector encoding, we examined the variance of HOP distance under the $10$ random orders used in our experiments. 
Figure.~\ref{fig:HOP_mean_std} shows the mean and standard deviation of the normalized HOP distance over the $10$ random taxa orders.
We can observe that taking the mean HOP distance over 10 random taxa orders results in a variation over the $50$ ytrees comparable to what we observe with the RF distance.
Table~\ref{tab:correlation} shows that the mean HOP distance correlates much better with the actual number of SPR steps that th RF distance.
Given that the HOP distance is very fast to compute, this suggests that applying it with a small number of random taxa orders allows to counterbalance the impact of a chosen taxa order on the distance value, at a minimal computational cost.

\begin{table}[htbp]
    \begin{center}
        \begin{tabular}{l|ccccccccccc}
            \hline
            $k$  &  10 & 20 & 30 & 40 & 50 & 60 & 70 & 80 & 90 & 100 \\
            \hline\hline
            RF       &0.815 &0.902 &0.909 &0.913 &0.909 &0.905 &0.900 &0.892 &0.883 &0.874\\
            HOP      &0.707 &0.867 &0.909 &0.924 &0.928 &0.928 &0.926 &0.921 &0.916 &0.910\\
            Mean HOP &0.840 &0.935 &0.953 &0.956 &0.954 &0.950 &0.944 &0.938 &0.931 &0.923\\
            \hline
        \end{tabular}
        \caption{Pearson correlation between the number of SPR steps and the RF and HOP distances. Every column corresponds to all comparison between the starting tree and the trees obtained by at most $k$ SPR steps. RF: Robinson-Foulds distance. HOP: HOP distance cumulated over the 10 random taxa orders. Mean HOP: mean of the HOP distance over the 10 random taxa orders.}
        \label{tab:correlation}
    \end{center} 
\end{table}

\paragraph{HOP distance reveals useful information on tree similarity.}
In Figure~\ref{figS1_2trees}, two phylogenetic trees depicting $91$ fungus species are presented. 
These trees were constructed by rooting on the branch leading to {\it Aspergillus aculeatus}, derived from two unrooted trees generated using IQTREE with different parameters, given  in \cite{shen2020investigation}. 
The HOP distance between these two trees is $10$, with the species ordered as they appear in the Newick representation of the first tree. 
Following Proposition~\ref{prop:HOP_forest}, the LCS defined by the HOP distance defined a forest of three distinct large subtrees (Fig.~\ref{figs2_commontrees}), illustrating a case where a moderate HOP distance translates into large conserved structure.

%% file: conclusion.tex
\section{Discussion and Conclusion}
\label{sec:conclusion}

We have introduced a novel representation of phylogenetic trees (with polytomies allowed) as vectors. 
This representation is slightly more efficient than the Newick format to store trees. 
In this representation, the internal nodes of a tree are represented by taxa and listed in a topological sorting ordering. 
This ordering ensures that the orientation of tree branches aligns with the sequence of internal nodes in the representation. 
Accordingly, as detailed in the Appendix, this representation enables us to develop an alternative method for counting phylogenetic trees and an algorithm for verifying tree identities through direct comparison of their vector representations.
It also leads to an efficient algorithm for computing the RF distance.  
Furthermore, since this vector representation can be extended to tree-child networks, exploring its applications in the network domain presents an intriguing avenue for future research.

We also introduce the HOP rearrangement operator and the HOP distance metric. 
The HOP rearrangement is implemented by moving a single entry in the vector representation of a tree, and is thus easy to implement. 
Although the HOP operator is a special type of SPR, the neighbourhood size for it remains quadratic in order and  the maximum HOP distance between two trees is less then $n$ HOP rearrangements.
Unlike most tree rearrangement operators, such as NNI and SPR, the HOP distance can be computed efficiently, in near-linear time. 
Moreover, given two trees, it is easy to compute a \textit{shortest} sequence of HOP rearrangements that transform a tree into the other.
Our experiments suggest that the HOP distance captures the dissimilarity between trees better than the RF distance.
Most importantly, the computation of the transformation between two trees allows us to identify the common subtrees shared by the trees, even with thousands of taxa. 

One feature of the HOP distance is its dependence on the taxon order used to encode trees into vectors. 
This characteristic is inherent to many dimensionality reduction techniques in data science, where the focus is on finding a good solution rather than the best one to overcome the challenge of large datasets. 
For efficient storage, the representation under any taxon order serves the purpose well. 
For the measurement of the dissimilarity between trees, our experiments show that the variation of HOP distance from different taxa orders is significant, particularly for very similar trees.
Consequently, the HOP distance can serve as a viable proxy for the SPR distance, especially for large trees where computing the SPR distance becomes impractical. 
In cases where the SPR distance approaches the number of taxa, the HOP distance is expected to provide a reliable measure of dissimilarity. 
Furthermore, for similar trees, the use of the mHOP distance, which minimizes the HOP distance over all possible taxon orders, is recommended. While the NP-hardness of computing the mHOP distance remains unknown, it is plausible to devise a fast algorithm for it by sampling taxon orderings, as demonstrated in~\cite{zhang2023fast}  for inferring phylogenetic networks.

Lastly, due to the efficiency of our vector representation in storage and the ease of implementing the HOP operator, there is potential for accelerating the development of machine learning methods for inferring phylogenetic trees. 
Exploring this research direction holds significant promise.

%% file: appendix.tex
\section*{Appendix: Additional results}

\subsection*{Derivation of the counting formula}

Each phylogenetic tree has a unique vector representation given a fixed taxa ordering. 
Consider a tree vector $\mathbf{v}$ on $X=\{1, 2, \cdots, n\}$. 
Since the second copy of $n$ follows right after the second copy of $n-1$ in $\textbf{s}$, the removal of the first and second copy of $n$ results in a tree vector on $X'=X\setminus \{n\}=\{1, 2, \cdots, n-1\}$. Conversely, inserting the first copy of $n$ to any of the $2n-3$ positions between the first entry (which is $1$) and the last entry and appending the second copy of $n$ a the end of a tree representation on $X'$, we obtain a unique tree vector on $X$.

Taken together, the above two facts imply that the number of phylogenetic trees  on $X$ is as many as $2n-3$ times that on $X'$.
Consequently, there are $1\times 3\times \cdots \times (2n-3)=(2n-3)!!$ phylogenetic trees on $X$.

\subsection*{Identity verification for phylogenetic trees}

Since each tree has a unique tree representation for a given taxa ordering, one can check whether two trees are identical or not by direct comparison of their representation vectors under a specific ordering. 
This leads to a simple linear time algorithm for the isomorphism problem for phylogenetic trees.

\subsection*{Computing the clusters and the RF distances}


The RF distance is one of the most popular metrics for measuring the dissimilarity  between two phylogenetic trees. 
Recall that the RF distance between two phylogenetic trees is defined as the number of clades appearing in one tree but not in the other. 
Here, we introduce a new algorithm for computing the RF distance between two phylogenetic trees. 
It takes 2$n$ set operations at most for comparing two $n$-taxa phylogenetic trees.

According to the decoding algorithm, for each edge $(u, v)$ of a phylogenetic tree $T$, $u$ always appears before the $v$ in the representation $\mathbf{v}(T)$. 
Therefore, we can use the corresponding tree vector $\mathbf{v}(T)$ to compute the clades of the trees in linear time as follows:
\begin{itemize}
    \item Let $T$ contain $n$ taxa. 
    We use an $n+1$-dimensional array $A_v=(a_0, a_1, \cdots, a_n)$ to represent the clade for an internal node $v$, where $a_0$ is the number of the taxa in the clade  and for $i>0$, $a_i$ is $1$ if taxon $i$ is in the clade and is $0$ otherwise. 
    \item Let $\mathbf{v}(T)=(v_1,v_2,\dots, v_{2n})$. 
    We use $c(i)$ to denote the clade of the internal node represented by $v_i$.  
    If $v_i$ is the first copy of a taxon, then, $c(i)$ is the union of $c(i+1)$ and $c(k)$, where $k$ is the index such that $v_{k-1}$ is the second  copy of taxon $v_i-1$. 
    \item In this way, we can compute all the $n$ multiple-taxa clades and their cardinality in $n$ set operations and $n$ additions. 
\end{itemize}

Recall that in the representation of a tree, each internal node is represented by the first occurrence of a taxon symbol and appears in the LTS of a unique taxon. 

\begin{proposition}
\label{prop:RF1}
    Let $T_1$ and $T_2$ be two phylogenetic trees on the same set of taxa.

    (a) Let $t_1t_2\cdots t_k$ be the LTS of a taxon in a tree. 
    Then, the clades of the nodes represented by $t_i$ satisfies that 
    $\vert c(t_i) \vert > \vert c(t_{i+1})\vert $ for each $i\leq k-1$.
    
    (b) For any internal nodes $u_1\in V(T_1)$ and $u_2\in V(T_2)$,  they are on the LTS of the same taxon in the trees if $c_{T_1}(u_1)=c_{T_2}(u_2)$.
\end{proposition}

\medskip
\noindent \textbf{Proof.}  (a) The property is derived from  the node represented by $t_i$ being the parent of the node represented by $t_{i+1}$ for each possible $i$.

(b)  It is clear that only the clades defined by the nodes in the unique path from the root to the leaf $\ell$ contains $\ell$. 
Recall that $\min(S)$ denotes the smallest taxon for a taxon subset $S$.  
Therefore,  by definition, for any taxon $t < \min_{T_1}(u_1)$, $u_1$ is not in the LTS of $t$. 
On the other hand, $u_1$ is on the path from root to the leaf labeled with taxon $\min_{T_1}(u_1)$. 
If  $c_{T_1}(u_1)=c_{T_2}(u_2)$, we assume $t$ is their smallest taxon. 
Then $u_1$ and $u_2$ appear in the LTS of $t$  in $T_1$ and $T_2$ under the indexing used, respectively. 
This concludes the proof.
\\

For a taxon $t$ in a tree $T$, we use $\mathrm{LTS}_T(t)$ to denote the LTS of $t$ in $T$. 
By Proposition~\ref{prop:RF1}.(a), the clades of the nodes appearing in each LTS satisfy the inclusion property, that is that if $u_1$ appears before $u_1$ in an LTS, $c(u_1)\supset c(u_2)$. 
By  Proposition~\ref{prop:RF1}.(b), if  two clades are equal, the corresponding nodes must be in the LTS of the same taxa. 
Therefore, we can compute the number of common clades for two trees by comparing only the clades appearing in the LTS of a taxon in one tree with those appearing in the LTS of the same taxon in the other tree. 
This approach for computing the RF-distance is summarized in the algorithm below.
\\
\\
\begin{tabular}{l}
\hline
 {\sc RF-Distance Algorithm}\\
 \textbf{Input} The vector representations of $T_1$ and $T_2$ with annotated node clades\\
 \textbf{Output} The RF distance $d$ between $T_1$ and $T_2$\\
 \\
 1. $d = 0;$\\
 2. For each taxon $t$:\\
 \hspace*{1em} 2.1 $u_1u_2\cdots u_{p} \leftarrow \mathrm{LTS}_{T_1}(t) $; \;\;$v_1v_2\cdots v_{q} \leftarrow \mathrm{LTS}_{T_2}(t)$\\
\hspace*{1em} 2.2. $i =1$; $j=1$; \\
\hspace*{1em} 2.3. While ($i\leq p$ and $j\leq q$):\\
\hspace*{2em} 2.3.1. If $\vert c(u_i)\vert > 
      \vert c(v_j)\vert $: \{ $d = d+1$; $i = i+1$; \}\\
\hspace*{2em} 2.3.2. If $\vert c(u_i)\vert < 
      \vert c(v_j)\vert $: \{ $d = d+1$; $j = j+1$; \}\\  
\hspace*{2em} 2.3.3. If $\vert c(u_i)\vert = 
      \vert c(v_j)\vert $: \{if $ c(u_i) \neq   c(v_j)$: \{$d = d+2$; $i= i+1$; $j = j+1$;\} \}\\
3. Output d;\\
 \hline
\end{tabular}

\newpage

\setcounter{figure}{0}
\renewcommand{\figurename}{Figure}
\renewcommand{\thefigure}{S\arabic{figure}}

\section*{Supplementary Figures}
\thispagestyle{empty}

\begin{figure}[b!]
    \centering
\includegraphics[scale=0.4]{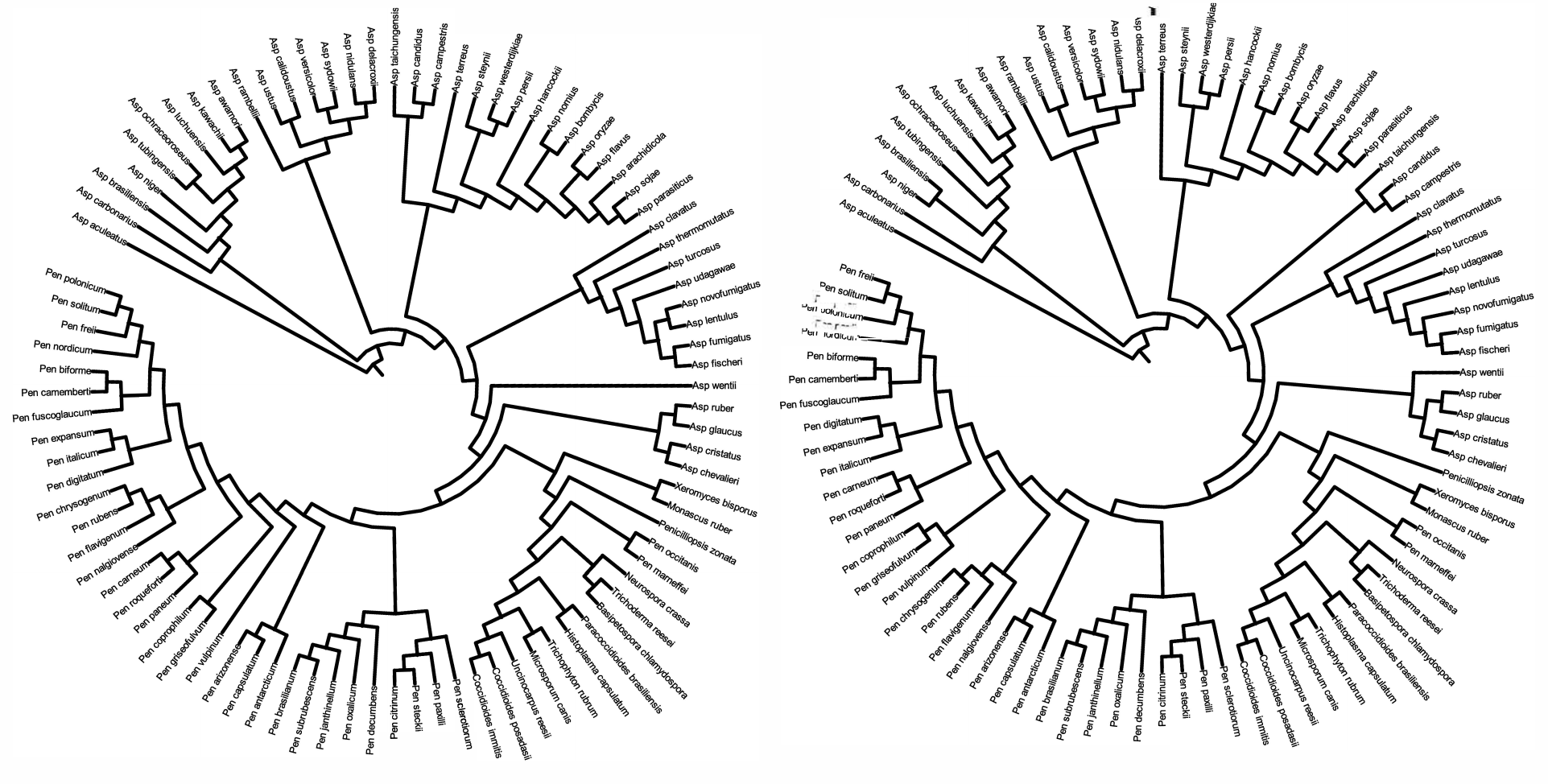}
\caption{Two phylogenetic trees inferred using IQtree with different parameters on 44 Aspergillus, 35 Penicillium and 12 closely related fungus species, appearing in \cite{shen2020investigation}.
\label{figS1_2trees}
}
\end{figure}

\begin{figure}[b!]
    \centering
\includegraphics[scale=0.6]{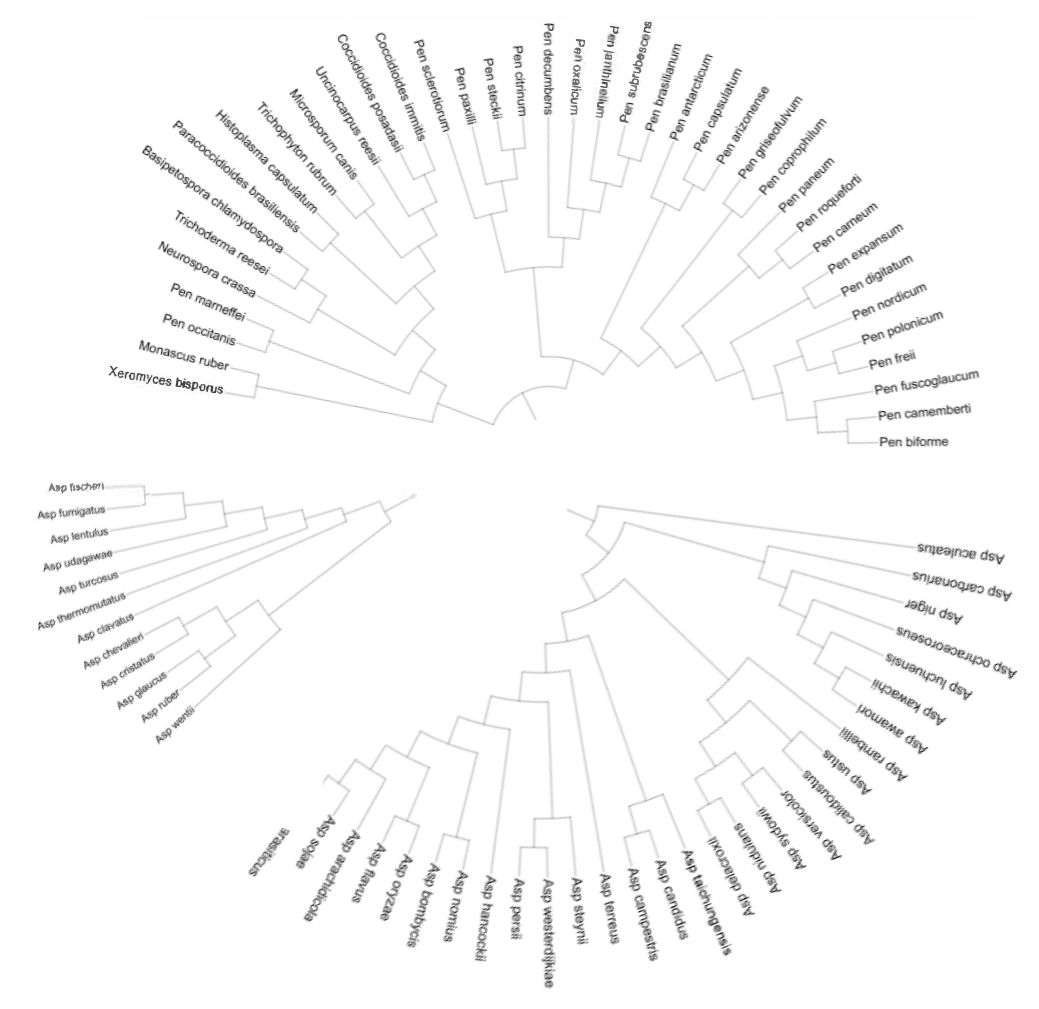}
\caption{The three largest common subtrees revealed by the HOP distance of 10 for the two trees in Fig~\ref{figS1_2trees}. The largest subtree contains 29  species. The other two contain respectively 29 and 12 Aspergillus species. \label{figs2_commontrees}
}
\end{figure}

%% file: main.bbl
\begin{thebibliography}{99}

\bibitem{felsenstein2004inferring}
Felsenstein J. 2004 {\em Inferring phylogenies}.
Sunderland, MA., USA: Sinauer Associates.

\bibitem{dayhoff1975evolution}
Dayhoff MO, McLaughlin PJ, Barker WC, Hunt LT. 1975  Evolution of sequences within protein superfamilies. {\em {Die Naturwissenschaften}} \textbf{62}, 154--161.

\bibitem{murphy2005dynamics}
Murphy WJ, Larkin DM, der Wind AEv, Bourque G, Tesler G, Auvil L, Beever JE, Chowdhary BP, Galibert F, Gatzke L et~al.. 2005  Dynamics of mammalian chromosome evolution inferred from multispecies comparative maps. {\em Science} \textbf{309}, 613--617.

\bibitem{redding2008evolutionarily}
Redding DW, Hartmann K, Mimoto A, Bokal D, DeVos M, Mooers A{\O}. 2008  Evolutionarily distinctive species often capture more phylogenetic diversity than expected. {\em {Journal of Theoretical Biology}} \textbf{251}, 606--615.

\bibitem{cann1987mitochondrial}
Cann RL, Stoneking M, Wilson AC. 1987  Mitochondrial {DNA} and human evolution. {\em Nature} \textbf{325}, 31--36.

\bibitem{ingman2000mitochondrial}
Ingman M, Kaessmann H, P{\"a}{\"a}bo S, Gyllensten U. 2000  Mitochondrial genome variation and the origin of modern humans. {\em Nature} \textbf{408}, 708--713.

\bibitem{boni2020evolutionary}
Boni MF, Lemey P, Jiang X, Lam TTY, Perry BW, Castoe TA, Rambaut A, Robertson DL. 2020  Evolutionary origins of the {SARS-CoV-2} sarbecovirus lineage responsible for the {COVID-19} pandemic. {\em {Nature Microbiology}} \textbf{5}, 1408--1417.

\bibitem{sharp2010evolution}
Sharp PM, Hahn BH. 2010  The evolution of {HIV-1} and the origin of {AIDS}. {\em {Philosophical Transactions of the Royal Society B: Biological Sciences}} \textbf{365}, 2487--2494.

\bibitem{foulds1982steiner}
Foulds LR, Graham RL. 1982  The {S}teiner problem in phylogeny is {NP}-complete. {\em {Advances in Applied mathematics}} \textbf{3}, 43--49.

\bibitem{roch2006}
Roch S. 2006  A short proof that phylogenetic tree reconstruction by maximum likelihood is hard. {\em {IEEE/ACM Transactions on Computational Biology and Bioinformatics}} \textbf{3}, 92--94.

\bibitem{stamatakis2005raxml}
Stamatakis A, Ludwig T, Meier H. 2005  {RAxML-III}: a fast program for maximum likelihood-based inference of large phylogenetic trees. {\em Bioinformatics} \textbf{21}, 456--463.

\bibitem{minh2020}
Minh BQ, Schmidt HA, Chernomor O, Schrempf D, Woodhams MD, Von~Haeseler A, Lanfear R. 2020  {IQ-TREE} 2: new models and efficient methods for phylogenetic inference in the genomic era. {\em {Molecular Biology and Evolution}} \textbf{37}, 1530--1534.

\bibitem{suchard2018bayesian}
Suchard MA, Lemey P, Baele G, Ayres DL, Drummond AJ, Rambaut A. 2018  Bayesian phylogenetic and phylodynamic data integration using {BEAST} 1.10. {\em {Virus Evolution}} \textbf{4}, vey016.

\bibitem{voznica2022deep}
Voznica J, Zhukova A, Boskova V, Saulnier E, Lemoine F, Moslonka-Lefebvre M, Gascuel O. 2022  Deep learning from phylogenies to uncover the epidemiological dynamics of outbreaks. {\em {Nature Communications}} \textbf{13}, 3896.

\bibitem{kim2020distance}
Kim J, Rosenberg NA, Palacios JA. 2020  Distance metrics for ranked evolutionary trees. {\em {Proceedings of the National Academy of Sciences U.S.A.}} \textbf{117}, 28876--28886.

\bibitem{penn2023phylo2vec}
Penn MJ, Scheidwasser N, Khurana MP, Duch{\^e}ne DA, Donnelly CA, Bhatt S. 2023  {Phylo2Vec}: a vector representation for binary trees. {\em arXiv preprint arXiv:2304.12693}.

\bibitem{liu2022analyzing}
Liu P, Biller P, Gould M, Colijn C. 2022  Analyzing phylogenetic trees with a tree lattice coordinate system and a graph polynomial. {\em {Systematic Biology}} \textbf{71}, 1378--1390.

\bibitem{prufer1918neuer}
Pr\"{u}fer H. 1918  Neuer beweis eines satzes \"{u}ber per mutationen. {\em {Archiv der Mathematik und Physik}} \textbf{27}, 742--744.

\bibitem{rohlf1983numbering}
Rohlf JF. 1983  Numbering binary trees with labeled terminal vertices. {\em {Bulletin of Mathematical Biology}} \textbf{45}, 33--40.

\bibitem{moore1973iterative}
Moore GW, Goodman M, Barnabas J. 1973  An iterative approach from the standpoint of the additive hypothesis to the dendrogram problem posed by molecular data sets. {\em {Journal of Theoretical Biology}} \textbf{38}, 423--457.

\bibitem{dasgupta1997distance}
DasGupta B, He X, Jiang T, Li M, Tromp J, Zhang L. 1997  On distances between phylogenetic trees. In {\em {ACM-SIAM Symposium on Discrete Algorithms}} pp. 427--436.

\bibitem{colienne2021computing}
Colienne L, Gavryushkin A. 2021  Computing nearest neighbour interchange distances between ranked phylogenetic trees. {\em {Journal of Mathematical Biology}} \textbf{82}, 8.

\bibitem{robinson1981}
Robinson DF, Foulds LR. 1981  Comparison of phylogenetic trees. {\em {Mathematical Biosciences}} \textbf{53}, 131--147.

\bibitem{zhang2023fast}
Zhang L, Abhari N, Colijn C, Wu Y. 2023  A fast and scalable method for inferring phylogenetic networks from trees by aligning lineage taxon strings. {\em {Genome Research}} \textbf{33}, 1053--1060.

\bibitem{cardona2008comparison}
Cardona G, Rossell{\'o} F, Valiente G. 2008  Comparison of tree-child phylogenetic networks. {\em {IEEE/ACM Transactions on Computational Biology and Bioinformatics}} \textbf{6}, 552--569.

\bibitem{stjohn2016review}
St.~John K. 2016  {Review Paper: The Shape of Phylogenetic Treespace}. {\em {Systematic Biology}} \textbf{66}, e83--e94.

\bibitem{allen2001subtree}
Allen BL, Steel M. 2001  Subtree transfer operations and their induced metrics on evolutionary trees. {\em {Annals of Combinatorics}} \textbf{5}, 1--15.

\bibitem{li1996nearest}
Li M, Tromp J, Zhang L. 1996  On the nearest neighbour interchange distance between evolutionary trees. {\em {Journal of Theoretical Biology}} \textbf{182}, 463--467.

\bibitem{hunt1977fast}
Hunt JW, Szymanski TG. 1977  A fast algorithm for computing longest common subsequences. {\em {Communications of the ACM}} \textbf{20}, 350--353.

\bibitem{shen2020investigation}
Shen XX, Li Y, Hittinger CT, Chen XX, Rokas A. 2020  An investigation of irreproducibility in maximum likelihood phylogenetic inference. {\em Nature communications} \textbf{11}, 6096.

\end{thebibliography}
